# High-throughput screening of metal-porphyrin-like graphenes for selective capture of carbon dioxide

Hyeonhu Bae[1,†], Minwoo Park[1,†], Byungryul Jang[1], Yura Kang[2], Jinwoo Park[2], Hosik Lee[3], Haegeun Chung[4], ChiHye Chung[5], Suklyun Hong[2], Yongkyung Kwon[1], Boris I. Yakobson[6], Hoonkyung Lee[1,*]

[1]School of Physics, Konkuk University, Seoul 143-701, Korea

[2]Department of Physics and Graphene Research Institute, Sejong University, Seoul 143-747, Korea

[3]School of Mechanical and Advanced Materials Engineering, Ulsan National Institute of Science and Technology, Ulsan 689-798, Korea

[4]Department of Environmental Engineering, Konkuk University, Seoul 143-701, Korea

[5]Department of Biological Sciences, Konkuk University, Seoul 143-701, Korea

[6]Department of Materials Science and Nanoengineering, Rice University, Houston, Texas 77005, United State

[†]These authors contributed equally to this work.

Number of pages: 24

**Correspondence to Hoonkyung Lee; correspondence and requests for materials should be addressed to H. L. (Email: hkiee3@konkuk.ac.kr).**

**Nanostructured materials, such as zeolites and metal-organic frameworks, have been considered to capture $CO_2$. However, their application has been limited largely because they exhibit poor selectivity for flue gases and low capture capacity under low pressures. We perform a high-throughput screening for selective $CO_2$ capture from flue gases by using first principles thermodynamics. We find that elements with empty *d* orbitals selectively attract $CO_2$ from gaseous mixtures under low $CO_2$ pressures (~$10^{-3}$ bar) at 300 K and release it at ~450 K. $CO_2$ binding to elements involves hybridization of the metal *d* orbitals with the $CO_2$ $\pi$ orbitals and $CO_2$-transition metal complexes were observed in experiments. This result allows us to perform high-throughput screening to discover novel promising $CO_2$ capture materials with empty *d* orbitals (e.g., Sc– or V–porphyrin-like graphene) and predict their capture performance under various conditions. Moreover, these findings provide physical insights into selective $CO_2$ capture and open a new path to explore $CO_2$ capture materials.**

Carbon dioxide gas is a greenhouse gas that is a primary cause of global warming, which is known to cause severe climate change[1]. In recent years, the temperature of the earth has increased because of significant increase in $CO_2$ emission. The emission of this gas is expected to continuously increase as the demand for fossil fuels increases, and thus the development of technologies for $CO_2$ capture is essential for addressing climate change[1]. The technology involving the capture of $CO_2$ gas from the flue gas is currently not sufficiently developed, particularly in the backdrop of the urgent need to reduce $CO_2$ emission.

Nanostructured materials, such as graphene, zeolites, and metal-organic frameworks, have been considered to capture $CO_2$. These materials are potentially useful because of their high capacity, fast $CO_2$ adsorption kinetics, and effective regeneration[2-11]. However, their application has been limited largely because they exhibit poor selectivity for flue gases and low capture capacity under low pressures (~$10^{-3}$ bar)[11-14], thereby limiting $CO_2$ capture from flue gases in power plants[14]. Thus, there is an increasing demand to search for novel $CO_2$ capture materials[15-17].

Recently, Fe–porphyrin-like fragments ($FeN_4$) to carbon nanotubes[18] and Co-porphyrin-like fragments ($CoN_4$) to nanostructures[19] were synthesized using the chemical vapour deposition and the pyrolysis methods, respectively, where Fe or Co is located at the center of four nitrogen atoms similar to metal-porphyrin structure[20,21]. We herein refer to this $MN_4$ structure as an M–porphyrin-like structure. Fused TM-porphyrin-like nanoclusters have been synthesized experimentally[22-26]. Furthermore, the porphyrin-like structure is analogous to the local structure of Fe in hemoglobin[27] or myoglobin[28], which deliver $O_2$ to the organs in the body. The concentration of nitrogen in carbon nanotubes and graphene has been found to reach ~8%[29] and ~10%[30], respectively. Thus, we expect that TM–porphyrin-like nanostructures can be synthesized experimentally. In this article, we perform first-principles thermodynamics based high-throughput screening for suitable M elements as selective $CO_2$ attractors using M–porphyrin-like graphene.

**Results**

To measure the $CO_2$ capture capabilities of nanomaterials from a mixed gas, we constructed a thermodynamic model of $CO_2$ adsorption on an adsorbent using the grand-canonical partition function[31]. We assumed a surface containing the number of identical, independent, and distinguishable adsorption sites ($N_s$) with no mixed adsorption of different molecules per adsorption site, wherein the number of adsorbed $i$-type gas molecules on the surface is $N_i$. If the adsorbed molecules and gases are in equilibrium, the grand partition function of the system can be written as

$$Z = (1 + \sum_i \sum_{n_i=1} g_{n_i} e^{n_i(\mu^i - \varepsilon^i_{n_i})/k_B T})^{N_s}, \qquad (1)$$

where superscript $i$ indicates the type of gas, $\mu^i (<0)$ denotes the chemical potential of the $i$-type gas, and $\varepsilon^i_{n_i} (<0)$ and $g_{n_i}$ denote the average adsorption energy and degeneracy of configuration (for a given adsorption number $n_i$) of the $i$-type gas molecules, respectively. When the thermally average number of $i$-type $CO_2$ is calculated from $<N_i> = k_B T \, \partial lnZ / \partial \mu^i$, the occupation function (i.e., coverage) of $CO_2$ for an adsorption site can be written as

$$f_{CO_2}(P,T) \equiv \frac{<N_{CO_2}>}{N_s} = \frac{\sum_{n_{CO_2}=1} n_{CO_2} g_{n_{CO_2}} e^{n_{CO_2}(\mu^{CO_2} - \varepsilon^{CO_2}_n)/k_B T}}{1 + \sum_i \sum_{n_i=1} g_{n_i} e^{n_i(\mu^i - \varepsilon^i_{n_i})/k_B T}}, \qquad (2)$$

Therefore, the thermodynamic $CO_2$ capture capacity of nanomaterials from a mixed gas can be computed using

$$C(P,T) = N_s f_{CO_2}(P,T) \Big/ \sum_i M_i m_i, \qquad (3)$$

where $M_i$ and $m_i$ denote the atomic mass and number of elements comprising the adsorbent, respectively.

The occupation function of $CO_2$ would have a positive value, i.e., $f_{CO_2} > 0$, if $\mu^{CO_2}(300K) > \varepsilon^{CO_2}$ and $\Delta^{CO_2} > \Delta^{other}$ at the adsorption (capture) conditions as shown in Figure 1a, wherein $\Delta^i \equiv \mu^i - \varepsilon^i$ is set and the superscript 'other' denotes molecules other than $CO_2$. In this case, selective $CO_2$ adsorption occurs through competitive adsorption between $CO_2$ and other molecules; this is attributed to the fact that the Gibbs factor for $CO_2$ adsorption is much greater than unity and the Gibbs factors of other molecules, i.e., $e^{(\mu^{CO_2}-\varepsilon^{CO_2})/k_BT} \gg 1$ and $e^{(\mu^{CO_2}-\varepsilon^{CO_2})/k_BT} \gg e^{(\mu^{other}-\varepsilon^{other})/k_BT}$. However, the occupation function would be zero, i.e., $f_{CO_2}=0$, if $\mu^{CO_2}(450K) < \varepsilon^{CO_2}$, at the desorption (release) conditions ($e^{(\mu^{CO_2}-\varepsilon^{CO_2})/k_BT} \ll 1$) as shown in Figure 1b, indicating that $CO_2$ adsorbed on the metal sites is released. Under a $CO_2$ pressure of $\sim 10^{-3}$ bar, the ideal conditions for adsorption and desorption are assumed to be 300 and 450 K, respectively, where $\mu^{CO_2}$ is approximately −0.75 and −1.20 eV, respectively, at ambient conditions. Thus, the key thermodynamic conditions for reversible and selective $CO_2$ capture from a mixed gas are as follows: (i) $-1.20\,\text{eV} < \varepsilon^{CO_2} < -0.75\,\text{eV}$ and (ii) $\Delta^{CO_2} > \Delta^{other}$.

From this we construct a computational approach to efficiently predict selective $CO_2$ capture materials based on first principles thermodynamics shown in Fig. 1(c). The thermodynamic conditions and capacity requirements[11] for screening are as follows: $-1.20\,\text{eV} < \varepsilon^{CO_2} < -0.75\,\text{eV}$ and $\Delta C(P,T) > 3\,\text{mmol g}^{-1}$ for $CO_2$ gas, and $\Delta^{CO_2} > \Delta^{other}$ and $\Delta C(P,T) > 3\,\text{mmol g}^{-1}$ for a mixed gas. $\Delta C(P,T)$ denotes the difference between $C(P,T)$ at 300 K and $C(P,T)$ at 450 K under a pressure of $10^{-3}$ bar, which indicates the $CO_2$ working capacity. These requirements may need to be revised depending on the operational environments.

We performed calculations on the adsorption energy of $CO_2$ molecules on the M sites of M–porphyrin-like graphene (Figure 2a). Elements of atomic numbers up to 92 for the M site were considered, and the others were ruled out because of their heavy weight. Sc–,

V–, Tc–, Os–, and Th–porphyrin-like graphenes out of many candidates met the reversibility requirements, viz. −1.2 to −0.8 eV (Figure 2a), where a $CO_2$ molecule adsorbs on a TM atom with the distance of ~2.5 Å between the TM atom and the $CO_2$ molecule. Therefore they were considered for the next step. We also performed $CO_2$ adsorption calculations on carbon allotropes such as carbon nanotubes, graphene, and $C_{60}$. The adsorption energy of the $CO_2$ molecule is ca. −0.05 eV, and the distance between their surface and the molecules is ~3.5 Å. In this case, since the adsorption energy of $CO_2$ molecules is much smaller than the required adsorption energy, pristine carbon nanostructures may not be suitable for use as $CO_2$ capture media under low pressure at room temperature. Notably, our approach significantly reduces the computational load because it is not necessary to calculate $\Delta C(P,T)$ for all the candidates in $CO_2$ gas or a mixed gas.

To predict the capture capabilities of the candidates, the $CO_2$ working capacities, $\Delta C(P,T)$, of the structures were computed using Eq. (3) (Figure 2b). The experimental values of the chemical potentials of $CO_2$ gas and calculated adsorption energies ($\varepsilon_n^{CO_2}$) were used in these calculations. Since the working capacities of Sc–, V–, and Tc–porphyrin-like graphenes satisfied the capacity requirement (>3 mmol g$^{-1}$), they were considered for the next selectivity screening step.

We observed three different geometries for the adsorbed $CO_2$ molecules on the TM atoms, which were designated as $\eta^1$-$CO_2$, $\eta^2$-$CO_2$, and $\eta^3$-$CO_2$, corresponding to the coordination numbers of the TM atom, i.e., 1, 2, and 3, respectively (Figure 3a). The adsorption energies of the $CO_2$ molecules were calculated to be −0.54, −0.79, and −0.78 eV per $CO_2$ for the Sc-$\eta^1$-$CO_2$, Sc-$\eta^2$-$CO_2$, and Sc-$\eta^3$-$CO_2$ geometries, respectively. The preferred $CO_2$ geometry depends on the metal type. The distance between the $CO_2$ molecule and TM atoms is 2.2–2.5 Å, which is much smaller than the equilibrium van der Waals distance (~3.4 Å), and the bond lengths of $CO_2$ are elongated by ~5%. Thus, the bonding between the TM atoms and $CO_2$ molecules must be chemical in nature.

To understand the enhanced interaction between early *d* orbital–containing elements and $CO_2$ molecules, we focused on a binding mechanism that appears between TM atoms and olefin molecules and is well known in organometallic chemistry[32]. The Dewar–Chatt–Duncanson model explains the type of chemical bonding between a π-orbital acid alkene and *d*-orbital metal atom by electron donation (i.e., hybridization of the empty *d* states with filled π states) and back-donation (i.e., hybridization of the filled *d* states with empty π states)[32]. The interaction between the TM *d* orbitals and the olefin π orbitals is called the "Dewar interaction". Therefore, empty *d*-orbital metals are expected to attract $CO_2$ molecules. The Dewar interaction is based on chemical bonding between the TM and $CO_2$ and can enhance the strength of the M–$CO_2$ bond beyond that of the van der Waals interaction. It is noteworthy that $Ca^{2+}$ also has empty 3*d* orbitals near the Fermi level that could participate in the Dewar interaction.

Next, we investigated whether the enhanced adsorption observed with early TM atoms is caused by the Dewar interaction. We observed the hybridization of the Sc 3*d* states with the $CO_2$ states at around −2.5, −2.0, and −2.0 eV for the $\eta^1$-$CO_2$, $\eta^2$-$CO_2$, and $\eta^3$-$CO_2$ geometries, respectively (Figure 3b). The difference in charge density between the Sc atom and $CO_2$ molecule (Figure 3c) indicates chemical bonding between $CO_2$ and the metal atoms. From this, we concluded that the enhanced binding of $CO_2$ to the metal atom originates from the Dewar interaction. The distinct adsorption geometries of $CO_2$ can be explained by the different hybridization states of the TM $d$ orbitals with the $CO_2$ $\pi$ orbitals (Figure 3d).

To examine the selectivity of $CO_2$ adsorption on Sc, V, and Tc sites in the presence of a mixed gas, we also carried out calculations on the adsorption of multiple $CO_2$ molecules or ambient gas molecules such as $N_2$, $CH_4$, and $H_2$ onto the metal atoms. Several $CO_2$, $H_2$, $N_2$, and $CH_4$ molecules bound to Sc, V, and Tc atoms (Figures. 4a and 4b, Figure 5). The difference between the chemical potential at 300 K and $10^{-3}$ bar and the adsorption energy of $CO_2$ (or other gas molecules) was calculated (Figure 4c) using experimental values of the chemical potentials of $CO_2$, $H_2$, $N_2$, and $CH_4$ gases. The chemical potentials of gases were obtained by fitting the experimental values to the following expression $\mu^i(P,T) = \mu^i_{ideal}(P,T) + A^i + B^i \times T$ where upper subscript $i$ indicates the type of gases, $\mu^i_{ideal}(P,T)$ denotes the chemical potential of an ideal monatomic $i$-type gas for a given the pressure $P$ and the temperature $T$, and $A^i$ and $B^i$ are fitted coefficients of $i$-type gas. The fitted coefficients are presented in Table 1. Since Sc and V, and not Tc, were found to satisfy the conditions for selective $CO_2$ adsorption ($\Delta^{CO_2} > \Delta^{other}$), they were considered for the next screening step

We also considered the zero-point vibrational energy of the gas molecules adsorbed onto the TM atoms. This energy was calculated to be in the order of a few meV regardless of the metal. Since the zero-point vibrational energy is negligible compared to the (static) adsorption energy (Figure 4a), we ignored the influence of the zero-point vibrational energy on adsorption in all cases except for $H_2$. Since the zero-point energy of the $H_2$ molecules adsorbed on TM atoms was not negligible (25% of the calculated values), we corrected the $H_2$ adsorption energies to determine the true adsorption energy.

The statistical model obtained here can correctly describe the adsorption of $CO_2$ onto TM–porphyrin-like graphene in the presence of a mixed gas because the mixed adsorption of different molecules onto a TM atom is not energetically favorable. For instance, the adsorption energy at which both a $CO_2$ and $N_2$ molecule adsorb onto a Sc atom was calculated to be −0.9 eV, which is much higher than that (−1.3 eV) at which single $CO_2$ or $N_2$ molecules adsorb on different sites.

The $CO_2$ capture capacities, $C(P,T)$, from mixed gases with different compositions were calculated for Sc– and V–porphyrin-like graphenes (Figures. 6a and 6b). The ratios of the mixed gases were based on experimental measurements[4,33] from pre-combustion, post-combustion, and oxyfuel-combustion $CO_2$ capture. These results show high $CO_2$ selectivity of Sc– and V–porphyrin-like graphene in mixed gases, which is consistent with the prediction of the selectivity requirement of $\Delta^{CO_2} > \Delta^{other}$. The $CO_2$ working capacities, $\Delta C(P,T)$, of Sc– and V–porphyrin-like graphenes can reach ~4 mmol $g^{-1}$ (Figures 6c and 6d), which meets the capacity requirement of 3 mmol $g^{-1}$ in a mixed gas. Therefore, Sc– and V–porphyrin-like graphene were found to be suitable for highly selective $CO_2$ capture from flue gases at ambient conditions. Furthermore, the $CO_2$ pressure range covers the pressure (~$0.4 \times 10^{-3}$ bar) of $CO_2$ in the atmosphere because the concentration of $CO_2$ in the atmosphere is ~400 ppm.

**Discussion**

We performed first-principles total energy calculations regarding $CO_2$ adsorption onto metal–porphyrin-like structures to explore the feasibility of achieving room-temperature $CO_2$ capture under low pressures. We found that transition metal–porphyrin-like structures adsorb $CO_2$ molecules with the desirable binding energy range and the practical (usable) capacity under ambient conditions can reach ~3 mmol/g. Equilibrium thermodynamics studies showed that Sc– or V–porphyrin-like graphene structures were found to be suitable for use as room-temperature $CO_2$ capture media. These results indicate that nanostructures containing empty *d* orbitals may be applied for selective adsorption of $CO_2$ from flue gases. We believe our results provide a new approach to achieving $CO_2$ capture at room temperature.

We address the evidence of $CO_2$ binding to TM atoms for $CO_2$ capture. TM-$\eta^1$-$CO_2$ or TM-$\eta^2$-$CO_2$ complexes were observed in experiments[34,35]. The capture of $CO_2$ involved in the first step of carbon capture/storage (CCS) technology requires high energy consumption[36,37]. Thus, the development of media such as TM–porphyrin-like graphene nanostructures, which can selectively adsorb $CO_2$ at room temperature under low $CO_2$ partial pressure, is expected to lower the cost of $CO_2$ adsorption and make CCS more viable.

**Methods**

We performed first-principles calculations based on the density functional theory (DFT)[38] as implemented in the Vienna Ab-initio Simulation Package (VASP) with the projector augmented wave (PAW) method[39]. The generalized gradient approximation (GGA) in the Perdew–Burke–Ernzerhof scheme[40] was used for the exchange correlation energy functional, and the kinetic energy cutoff was taken to be 800 eV. For calculations of gas molecule adsorption, our model for the graphene-based system comprised a 3 × 3 hexagonal supercell, and the composition of the supercell was $C_{12}N_4M_1$. Geometrical optimization of the graphene-based system was carried out until the Hellmann–Feynman force acting on each atom was less than 0.01 eV/Å. The first Brillouin zone integration was performed using the Monkhorst–Pack scheme[41]. 4 × 4 k-point sampling was used for the 3 × 3 graphene supercells. The chemical potential of gases, $\mu = (H - TS)/N$, where *H*, *S*, and *N* denote the enthalpy, the entropy, and the number of particles was calculated from the data of the enthalpy (*H*) and entropy (*S*) in the reference: http://webbook.nist.gov/chemistry/fluid/.

**Acknowledgments**

We thank C.-H. Park for critical reading. This research was supported by WTU Joint Research Grants of Konkuk University, the Basic Science Research Program (H. L.[a]: 2015R1A1A1A05001583), and Nano·Material Technology Development Program (S. H.: 2012M3A7B4049888) through the National Research Foundation of Korea (NRF) funded by the Ministry of Science, ICT & Future Planning. The Priority Research Center Program (S. H.:2010-0020207) through NRF funded by the Ministry of Education (MOE) also supported this work.

**Author Contributions**

H. B. and M. P. contributed equally to this work. H. L.[a] conceived and designed the study. H. B., M. P., B. J., and J. P. performed the calculations. Y. K.[a], H. L.[b], H. C., C. C., S. H., Y. K.[b], B. I. Y., and H. L.[a] interpreted the data. C. C. and H. L.[a] wrote the manuscript. All authors revised the manuscript and approved the final version of the manuscript.

**Author Information**

Supplementary information is available in the online version of the paper. Reprints and permissions information is available at www.nature.com/reprints. Correspondence and requests for materials should be addressed to H. L.[a] (hkiee3@konkuk.ac.kr).

**Competing Financial Interests**

The authors declare no competing financial interests.

**Table 1.** The fitted chemical potentials of gases.

| Gas type | $A^i$ (eV) | $B^i$ (meV/K) | $R^2$ |
|---|---|---|---|
| $CO_2$ | 0.04271 | −0.6425 | 0.99882 |
| $H_2$ | 0.02784 | −0.1585 | 0.99590 |
| $N_2$ | 0.03000 | −0.4512 | 0.99948 |
| $CH_4$ | 0.04868 | −0.4840 | 0.99823 |

$R^2$ is the coefficient of determination (measure of goodness of fit).

**Figures legends**

**Figure 1.** Thermodynamics of reversible/selective adsorption of $CO_2$ and flow chart for predicting selective $CO_2$ capture materials: (a) Selective $CO_2$ adsorption occurs through competitive adsorption between $CO_2$ and other molecules if $\mu^{CO_2}(300\text{ K}) > \varepsilon^{CO_2}$ and $\Delta^{CO_2} > \Delta^{other}$. (b) $CO_2$ molecules adsorbed on the metal sites are released if $\mu^{CO_2}(450\text{ K}) < \varepsilon^{CO_2}$. (c) Flow chart for predicting reversible and selective $CO_2$ capture materials based on first principles thermodynamics: this consists of reversibility screening for pure $CO_2$ gas and selectivity screening for a mixed gas.

**Figure 2.** Reversibility screening of many candidates: (a) Calculated adsorption energies of $CO_2$ molecules on M–porphyrin-like graphene and a variety of nanostructures. Inset shows the schematic of $CO_2$ binding to the M site and colored-marked elements indicate data not available. (b) Calculated $CO_2$ capture capacity, $C(P,T)$, on M–porphyrin-like graphene for $CO_2$ gas at 300 K under a $CO_2$ pressure of $10^{-3}$ bar. Colored-marked bars indicate candidates which meet the requirements.

**Figure 3.** Origin of distinct geometries of $CO_2$ adsorption: **(a)** Atomic structures showing $CO_2$ molecule adsorbed onto Sc–4N graphene for the various $CO_2$ adsorption geometries designated as $\eta^1$-$CO_2$, $\eta^2$-$CO_2$, and $\eta^3$-$CO_2$, respectively. **(b)** The density of states for $\eta^1$, $\eta^2$, and $\eta^3$ geometries, respectively. **(c)** The difference in the total charge density $\Delta\rho = \rho(GP+4N+Sc+CO_2) - \rho(GP+4N+Sc) - \rho(CO_2)$ for $\eta^1$, $\eta^2$, and $\eta^3$ geometries, respectively. Yellow and green indicates the charge accumulation and depletion. **(d)** The schematic of the hybridization of the Sc $3d$ orbitals with the $CO_2$ $p_z$ orbitals for $\eta^1$, $\eta^2$, and $\eta^3$ geometries, respectively. Red and blue colors of the orbitals indicate the different phases, respectively.

**Figure 4.** Selectivity screening by selective $CO_2$ capture condition: (a) Calculated (average) adsorption energies of molecules for the different types of molecules with different numbers of the molecules as TM atoms (TM = Sc, V, Tc). *, **, and *** indicate the geometric configurations of $\eta^1$, $\eta^2$, and $\eta^3$, respectively. (b) Optimized geometry of three $CO_2$ molecules adsorbed onto a Sc atom of Sc–porphyrin-like graphene with the $\eta^1$ configuration. (c) The difference ($\Delta^i = (\mu^i - \varepsilon^i_{n_i})n_i$) between the chemical potential of a gas and adsorption energy of the gas molecule on TM–porphyrin-like graphene with respect to the type of gas. The largest values of $\Delta^i$ were chosen regardless of $n_i$.

**Figure 5.** Adsorption of various molecules on Sc-porphyrin-like graphene: **(a)** Up to three $H_2$ molecules adsorb on a Sc atom. **(b)** Up to three $N_2$ molecules adsorb on a Sc atom. **(c)** Up to two $CH_4$ molecules adsorb on a Sc atom.

**Figure 6.** Selectivity screening by $CO_2$ working capacity: Calculated capacities, $C(P,T)$, of $CO_2$ of the TM–porphyrin-like graphenes as a function of temperature under total pressure, $P$, of $10^{-3}$ bar using Eq. (3): (a) Sc–porphyrin-like graphene and (b) V–porphyrin-like graphene. The following different compositions of gases were considered: Pure $CO_2$ (100%), $CO_2$ (89%)-$N_2$ (11%), $CO_2$ (40%)-$H_2$ (57%)-$N_2$ (3%), $CO_2$ (20%)-$H_2$ (75%)-$CH_4$ (5%), and $CO_2$ (17%)-$N_2$ (83%). The partial pressure of gases is given by $P_i = x_i P$, where $x_i$ is the composition of the gas. Calculated working capacities of $CO_2$ in the TM–porphyrin-like graphene as a function of the total pressure, $P$, of the gases from $\Delta C(P,T)$, the difference between $C(P,T)$ at 300 K and $C(P,T)$ at 450 K: (c) Sc–porphyrin-like graphene and (d) V–porphyrin-like graphene.